# Asteroid Resource Utilization: Ethical Concerns and Progress

**Lead Author:** Andrew S. Rivkin (Johns Hopkins University Applied Physics Laboratory, andy.rivkin@jhuapl.edu)

**Co-authors:** Moses Milazzo (Other Orb LLC; He/Him/His), Aparna Venkatesan (University of San Francisco), Elizabeth Frank (First Mode), Monica R. Vidaurri (Howard University/NASA Goddard Space Flight Center), Phil Metzger (Florida Space Institute/University of Central Florida), Chris Lewicki (Former CEO, Planetary Resources)

**Co-signers:** Barbara Cohen (NASA GSFC), Parvathy Prem (Johns Hopkins University Applied Physics Laboratory), Mark Gurwell (Center for Astrophysics | Harvard & Smithsonian), Cosette Gilmour (York University), Michael C. Nolan (University of Arizona), Divya M. Persaud (UCL), Alessondra Springmann (University of Arizona), Margaret Landis (University of Colorado, Boulder/LASP), Flaviane C. F. Venditti (Arecibo Observatory/UCF, Michele T. Bannister (University of Canterbury, NZ), M. M. McAdam (NASA Ames Research Center), Jennifer (JA) Grier (Planetary Science Institute), Katelyn Frizzell (Rutgers University), Alexandra Warren (University of Chicago), Amanda A. Sickafoose (Planetary Science Institute), Theodore Kareta (University of Arizona), Jamie Molaro (Planetary Science Institute), Anthony Hennig (George Washington University), Angela Stickle (Johns Hopkins University Applied Physics Laboratory)

**350-character abstract:**
**As asteroid mining moves toward reality, the high bar to entering the business may limit participation and increase inequality, reducing or eliminating any benefit gained by marginalized people or developing nations. Consideration of ethical issues is urgently needed, as well as participation in international, not merely multilateral, solutions.**

**Executive Summary**

The past decade has seen asteroid mining move from a science-fiction topic to a near-term likelihood. Several asteroid mining companies were founded in the last decade and developed candidate business cases. While most of these original companies have since been disbanded or moved into other business, it seems like it is only a matter of time until a well-funded (and sufficiently patient) effort moves forward, either as part of an effort to develop a market or in anticipation of one. Furthermore, unlike resources from the Moon or Mars, the main cases for use of asteroidal resources are focused on applications in Earth orbit (and perhaps eventually on Earth itself) and are largely if not entirely unrelated to human exploration, which in principle detaches asteroid mining development from the state of lunar or martian exploration.

This past decade has also seen a vast amount of wealth created through the development of technology. However, this wealth has largely been concentrated in the hands of a very small number of people, who as a result have achieved tremendous influence in Western society. It is not an exaggeration to note that everything from our elections to our pandemic response to our shopping and driving habits have been altered by platforms and technology that were in their infancy when the last Decadal Survey was underway.

The inequities inherent in access to space will lead to great disparities in who is able to take advantage of such a wealth of resources, and who benefits from (and who is hurt by) that exploitation. The most enthusiastic advocates of asteroid mining suggest it ultimately could unlock trillions of dollars of wealth, though more sober analysts think it may represent a few tens of billions of dollars over the next few decades at the low end. Regardless of amount, it is imperative that this wealth be used responsibly and for the good of all humankind. Ethical arguments as well as the history of wealth accumulation indicate that laws and regulations and the political will to enforce them will be necessary to ensure asteroid mining is a net gain for humanity. Ethical considerations must take an important place in making these laws and regulations, and we must act now to prevent harmful precedents from being set.

As space scientists, we are often told how much our work inspires future generations. That inspiration can cut both ways: how we act in taking these steps into the cosmos will set precedents that subsequent generations will either follow or have to work to undo. We have an opportunity and an obligation to influence careful consideration of not only *how* to make technical progress but *whether* we should, and how that progress can benefit everyone rather than just a few already-wealthy and powerful people. This white paper is being written during a pandemic, with climate change contributing to increased severity and frequency of extreme weather events according to the National Academies[1], and as the use of facial recognition software to aid arrests of protestors has increased anxiety about civil rights violations in the USA[2]. All of these issues have been affected directly or indirectly over short and long terms by ethical choices made by scientists, pointing to the importance of ethics at an individual level.

Several actions taken by private companies in space over the last several years make it obvious that there are at least some actors that cannot or will not self-regulate, do not consider the ethical ramifications of their actions, and often do not enter into conversations with all stakeholders until it is too late. The most striking of these actions are the launches of the first Starlink satellites, intended to be part of a megaconstellation with potentially dramatic negative impacts on our view of the night sky for observers ranging from casual stargazers to expensive international projects like the Vera Rubin Observatory (McDowell 2020, also see the report from the Satellite Constellations 1 Workshop[3] and the "Impact of Satellite Constellations" statement from the Rubin Observatory[4]). In the case of Starlink, the harm is ongoing and will potentially increase if mitigation plans for the next phase of the project fail or are discarded.

---

[1] National Academies story: https://bit.ly/2FwWz2r
[2] Story in Ars Technica: https://bit.ly/2ZyZTRA

In this white paper, some non-technical aspects of asteroid mining are addressed. We will note the ways in which important conversations are in the early stages or have not yet begun. We note a consensus that international agreements regarding asteroid mining are urgently needed to ensure the practice is conducted ethically and in accordance with international law, and we note that the US and especially NASA have been glaringly absent from recent conversations about the legal and ethical frameworks of In-Situ Resource Utilization (ISRU). We include recommendations where appropriate. We hope this white paper brings more awareness to these important issues both within and beyond the Decadal Survey and that it helps to guide the future, should asteroid mining become a part of the world economy as anticipated.

**Is it Ethical to Mine Asteroids?**

The Outer Space Treaty governs the acts of nations in space and forbids the claiming of territory. Each nation is responsible for the acts of companies incorporated in their territory. Interpretations of how the treaty applies to companies that seek to establish infrastructure and extract materials vary, leading several countries, notably Luxembourg, the USA, and the UAE, to establish laws intended to minimize or eliminate uncertainty and encourage investment. The consistencies of these national laws with international law have yet to be tested.

Separate from legal ramifications of ISRU, whether by national actors or individual or corporate actors are the ethical questions and implications of those actions. An action that is legal is not always ethical and it is important to distinguish the two.

The last several years have led many to look at the rationales for expanding human presence and the Earth's economy into space. In the past, many saw that expansion as our common "manifest destiny" and expected market forces to naturally lead that expansion. In an era of climate change, and a reexamination of American history by many, we might reasonably wonder what benefits asteroid mining will bring to the American people and the world, and whether public investments should be made to enable it. Pilchman (2015) notes that asteroid mining is likely to increase inequality on Earth, and thus be an unethical practice, unless it can be regulated to bring benefits to all[5]. Schwartz (2016) argued that mining of asteroid resources was unlikely to "significantly improve the well-being of average human beings" (and by extension, would be unethical), though he assumed those resources would be used to support space-based rather than Earth-based needs, a conclusion contradicted by Metzger (2016).

The planetary science community has begun reckoning with issues of diversity, equity, inclusion, and accessibility (DEIA) over the past decade in ways that it had not before, driven both by societal changes but also by issues more specific to the field including the siting of the Thirty Meter Telescope and the future of Mauna Kea Observatory. Venkatesan et al. (2019) discuss the need to integrate indigenous knowledge and mainstream astronomy to prevent the expansion of "the mindset of colonialism to a truly cosmic scale".

---

[3] https://aas.org/satellite-constellations-1-workshop-report, July 2020

[4] https://www.lsst.org/content/lsst-statement-regarding-increased-deployment-satellite-constellations, May 2020

[5] We note that while "fairness" and "equity" cannot be measured using SI units, social scientists, legal experts, and policymakers have documented and established metrics for comparing the fairness, equity, etc..

This "mindset of colonialism" is deeply intertwined with many stated motivations for resource exploitation in space and its use as a driver for human expansion into the Solar System. A white paper submitted by Tavares et al. (2020) argues that it is "critical that ethics and anticolonial practices are a central consideration of planetary protection". While planetary protection is the main focus of that white paper, it addresses several topics of concern here as well. The authors and co-signers recommend that the planetary science and space science community reevaluate the ethics of planetary missions in order to adequately address questions related to ISRU, among other topics, and to explore ethical questions that include the "preservation of environments on planetary bodies", the "long-term environmental impacts of resource extraction on planetary bodies", and the "short-term impact of largely unrestrained resource extraction on wealth inequality".

**Is it Ethical not to Mine Asteroids?**
On the flip side of the questions about whether it is ethical to mine asteroids is the question of whether it is ethical to leave a vast store of (probably) life-free resources untouched while continuing to extract resources from the only planetary body with known life in the Solar System. In the current paradigm of accelerating climate change and ecosystem collapse, it may become necessary to balance these two ethical weights against each other in a considered and thoughtful way. Given the UN paradigm discussed below that use of outer space is the "province of all" it is reasonable for society, which is being asked to fund investment in enabling technologies, to ask in return not only for a lack of harm from asteroid mining but for an equitable share of the positive benefits that will be gained.

**Current Technical Situation and the Potential Impacts on Ethical Considerations**
One possible benefit that has been mentioned is to use increased resource extraction from asteroids as a way to reduce resource extraction on Earth. Metzger (2016) argued that lowering the costs of infrastructure in space will allow solutions to Earth's increasing energy demands that are not currently feasible (such as beaming solar energy via microwave to Earth), and that lowering the cost of space infrastructure is reliant upon mining the asteroids and developing a space-based economy. The business models that have thus far been made public from extant or defunct asteroid mining companies all focus on the extraction and use of water or its constituent hydrogen and oxygen, and thus the near-term technical focus of research supporting asteroid mining focuses on the identification of water-rich near-Earth asteroids (NEAs) and the extraction of water (Elvis, 2014, Metzger 2017, Rivkin and DeMeo 2019).

The technical state of the art for asteroid mining, including knowledge gaps and near-term activities to bring asteroid mining closer to reality, is discussed in detail in a white paper by Lewicki et al. (2020). Demonstration missions were potentially only a few years from occurring and, presumably, could occur within a few years for interested and well-funded investors.

Many of these technical considerations should be considered in light of ethical concerns. Some specific legal issues are also timely: Wiegert (2020) studied the spread of ejecta from the DART impact into Dimorphos and concluded that while it was insignificant and posed no threat to Earth, repeated tests over long periods could generate a dust hazard in interplanetary space that was significant compared to the current background dust level. Repeated planetary defense tests like DART are unlikely, but Wiegert noted that mining operations could also plausibly generate such debris. Meteoroid stream formation via mining of large asteroids was considered by Fladeland et al. (2019), who found that the fluxes generated could be comparable to the natural meteoroid flux for sufficiently vigorous mining activity. The questions of whether a legal limit on ejected debris should exist and what that limit should be are not settled, though Fladeland et al. advocated for international solutions to this problem before a combination of lax oversight and a broad view of proprietary data lead to unsafe and unsustainable practice. There are likely dozens of similar issues that remain unexamined but that could and should lead to statutes and regulations on asteroid mining.

**Space Is The Common Heritage of Humanity**

In 1996, the United Nations adopted General Assembly Resolution 51/122. It noted that "International cooperation in the exploration and use of outer space for peaceful purposes…shall be the province of all mankind." The current administration in the United States has explicitly stated that it "does not view [outer space] as a global commons"[6], anticipating a series of bilateral or multilateral agreements between like-minded countries. Signers of a recent open letter to the UN General Assembly[7] expressed concern that such multilateral approaches "marginalize input from developing and non-spacefaring States", directly against the spirit of the 1996 resolution. The Hague International Space Resources Governance Working Group has developed a set of "building blocks" from which they hope an international framework may be built[8], with principles arising from the Outer Space Treaty and other UN statements and resolutions.

While several papers consider ethical issues in space science, those that we are aware of have addressed asteroid mining in broad terms and artificially siloed from some of the technical issues. The nature of asteroid mining provides unusual complications that may lead to unprecedented problems. While collection of space resources is occasionally compared to collection of resources from the sea[9], the barrier for entry to the latter is a fishing pole or net or a willingness to dive for pearls, while asteroid mining requires advanced technology and large amounts of starting capital. It also seems possible if not likely that the earliest successes in asteroid mining will be the only successes, as competition with established companies will provide an additional barrier, and a monopoly or cartel may develop. Legal steps must be taken to ensure such monopolies or cartels cannot exist or that humanity benefits even if they do develop. Given the possible timelines for asteroid mining, it is urgent to bring together relevant experts to discuss specific future issues that already are apparent.

**The Terrestrial Mining Experience**

Mining in particular has an infamous history linked to colonialism, with disastrous effects on indigenous populations around the world who were enslaved, forcibly relocated, and murdered to allow the redistribution of wealth from colonies to powerful empires. It is also one of humanity's oldest industries, and as such, we can turn to it for lessons and frameworks that might be applied to the future of asteroid mining.

A consequence of the negative impacts on human life and the environment caused by mining activities and their remnants (e.g., Cornwall, 2002) over the last few decades has been the creation of environmental and social regulations to minimize and mitigate the harms that mining can have on people and the environment. As described in detail in Vidaurri and Gilbert (2020), environmental impact assessments are a standard part of the process of regulatory approval for new large scale projects including but not limited to mines. They are mandated in the United States by the National Environmental Policy Act (NEPA), which requires federal agencies to mitigate impacts to the human environment posed by their work prior to its execution. Such a framework may be applicable to asteroid mining. Despite its deleterious impact on people and the environment, mining underpins modernity as we experience it in 2020, providing new materials for infrastructure, manufacturing, energy production (including green energy sources such as solar panels), large-scale agriculture, and more. Simultaneously, the general public's tolerance for these impacts has declined as their effects have accumulated, even as regulations have expanded. This legacy of bad decision-making throughout history has left a stain on the profession of mining, and planetary scientists would do well to avoid the same impacts to their profession. For example, the mining workforce is aging in part due to the challenge of attracting early career employees who are more environmentally minded than the previous generation (Human Resource Management, 2015).

---

[6] Executive Order on Encouraging International Support for the Recovery and Use of Space Resources, 6 April 2020.

[7] http://www.outerspaceinstitute.ca/docs/InternationalOpenLetterOnSpaceMining.pdf, August 2020.

[8] https://bit.ly/3bWTZP5, dated 12 November 2019, retrieved 2 September 2020.

[9] 'Bridenstine compared the approach to commercial fishing in international waters: "You do not own the ocean," he said, but "you own the tuna" you catch.': https://bit.ly/3hyILTh, reported September 10, 2020

Additionally, mining companies are increasingly concerned with obtaining "social license to operate" from local stakeholders, who will act to shut down mining operations (such as by striking or blocking mine entrances) if they are dissatisfied. Social license to operate is defined as the ongoing acceptance of a company or industry's standard business practices and operating procedures by its employees, stakeholders, and the general public. To obtain social license, beyond meeting any relevant federal and state regulations, mining companies must work to develop community relationships, though not always successfully: major projects will be canceled if social license to operate is not perceived to be obtained.

What does this mean for asteroid mining? No humans will be directly impacted through asteroid mining operations comparable to terrestrial mining with community displacement or disruption by noise and dust from mine operations. Nonetheless, all humans are arguably stakeholders in how the resources of the solar system are used for development, just as all humans are impacted by the growing number of Starlink satellites populating the night sky. Given that the barriers to asteroid mining tilt to the advantage of those who are already wealthy and well-capitalized, marginalized peoples and developing nations are likely to disproportionately suffer by exclusion from benefits. By acting outside the bounds or in the absence of multilateral space policy, asteroid mining companies are at risk of repeating the mistakes of their predecessors in pursuit of profit, particularly if they do not obtain social license to operate from affected stakeholders. It is not too soon for those stakeholders—and arguably all of humanity—to determine what obtaining that social license will require. It is also in the best interests of the would-be asteroid mining companies to be engaged with this process in good faith, lest they find themselves the recipients of backlash.

**Recommendations for the Next Decade**

Ethics papers that mention asteroid mining are in general agreement that it is only ethical if it leads to benefits for all of humanity, and that the only way to ensure that it will lead to such benefits is if there are appropriate laws and regulations in place (Pichman 2015, Schwartz and Milligan 2017). New laws have sought to clarify what is and is not permissible in terms of asteroidal and lunar resources, but we must move beyond what is merely permissible and also generate broad consensus on what is ethical. Given the pace of technology, the time to do this is now. This work should not be adversarial, nor seen as such, but is in the spirit of the view of the United Nations that space exploration should be done for the benefit of all.

1. **We recommend that experts in space science and engineering meet with ethicists, space lawyers, industry figures, and policy experts to discuss the specific ethical issues around asteroid mining and possible solutions. Because NASA, NSF, USGS, and other Federal agencies are developing technology and expertise that affect these questions, they should be involved.**
2. **We recommend that NASA, NSF, and other Federal agencies clarify the ways in which private, for-profit asteroid mining companies that use publicly-developed technologies (or financial incentives) will reward that public investment, with audits showing that those companies follow ethical guidelines as a condition of future funding. Given the high barriers to entry into asteroid mining and the risk of a future monopoly or cartels, a proactive approach to guaranteeing a share of public benefits is essential.**
3. **We recommend that within the US Dept of Commerce, a regulatory process step of an Environmental Impact Statement be made to raise awareness of the potential negative consequences of asteroid mining, along with actions already taken to mitigate those effects, mitigations not currently taken, and remaining items which can not be mitigated.**

Vidaurri et al. (2019) advocated the prioritization and improvement of ethics, planetary protection, and safety standards in the government and private sector as space exploration continues, arguing such priority must be executed not only by government policies but also by individual practice. While professional societies for space scientists have codes of conduct, those codes are understandably focused on interpersonal conduct and integrity of research. Neither explicitly considers questions of the larger ethical implications of specific research, save for experiments involving human or animal subjects.

4. **We recommend that professional societies in space sciences consider amending their codes of conduct to include an expectation that members will not undertake research or actions deemed to be unethical, even if human or animal subjects are not involved.**
5. **In conjunction with the amendment of codes of conduct, we also recommend that professional societies in the United States take up the cause of advocating for US leadership in developing international law and ethical standards that clarify the legal and ethical bounds of space mining.**
6. **We also urge individual scientists to consider their own ethical boundaries that they will not cross, even if legally allowed, and to do so in dialogue with colleagues, friends, employers, and ethicists.**

**Summary**

Developments over the last decade have led to a recognition that asteroid mining is a more realistic prospect than previously thought. Ethical and legal issues are major unsolved problems, and we recommend pursuit of international agreements to set standards that will bring maximum benefit to humankind as well as commitments by individuals to adhere to strict ethical standards as we seek to establish norms and precedents in asteroid mining practices.

**References:**

Cornwall, Warren (2020) Catastrophic failures raise alarm about dams containing muddy mine wastes. Science Magazine. https://bit.ly/3kfigCU.

Elvis, M. (2014). How many ore-bearing asteroids?. Planetary and Space Science, 91: 20-26.

Fladeland, L., Boley, A. C., & Byers, M. (2019). Meteoroid Stream Formation Due to the Extraction of Space Resources from Asteroids. *arXiv:1911.12840*.

Graps, A. L., et al.. (2016). ASIME 2016 White Paper: In-Space Utilisation of Asteroids:" Answers to Questions from the Asteroid Miners". *arXiv preprint arXiv:1612.00709*.

Graps, A. L., et al. (2019). ASIME 2018 White Paper. In-Space Utilisation of Asteroids: Asteroid Composition--Answers to Questions from the Asteroid Miners. *arXiv:1904.11831*.

Human Resource Management. (2015) Preparing for an Aging Workforce: Oil, Gas and Mining Industry Report.

Lewicki, C. et. al. (2020) Furthering Asteroid Resource Utilization in the Next Decade through Technology Leadership. *To be submitted as a Planetary Decadal White Paper.*

McDowell, J. C. (2020). The Low Earth Orbit Satellite Population and Impacts of the SpaceX Starlink Constellation. *ApJLett, submitted.* https://arxiv.org/abs/2003.07446

Mercer-Mapstone, L., Rifkin, W., Moffat, K., & Louis, W. (2017). Conceptualising the role of dialogue in social licence to operate. Resources Policy, 54, 137-146.

Metzger, P. T. (2016). Space development and space science together, an historic opportunity. *Space Policy*, *37*, 77-91.

Metzger, P. T. (2017). Economic Planetary Science in the 21st Century. In Planetary Science Vision 2050 Workshop, volume 1989 of LPI Contributions, page 8126.

Pilchman, D. (2015). Three Ethical Perspectives on Asteroid Mining. *Commercial Space Exploration: Ethics, Policy and Governance*, 135-147.

Rivkin, A. S., and F. E. DeMeo. How many hydrated NEOs are there?. Journal of Geophysical Research: Planets 124, (2019): 128-142.

Schwartz, J. S. (2016). Near-Earth water sources: Ethics and fairness. *Advances in Space Research*, *58*, 402-407.

Schwartz, J. S., & Milligan, T. (2017). Some ethical constraints on near-earth resource exploitation. In *Yearbook on Space Policy 2015* (pp. 227-239). Springer, Vienna.

Tavares, F. et al. (2020). Ethical Exploration and the Role of Planetary Protection in Disrupting Colonial Practices. *To be submitted as a Planetary Decadal White Paper.*

Venkatesan, A., et al. (2019). Towards inclusive practices with indigenous knowledge. *Nature Astronomy*, *3*, 1035-1037.

Vidaurri, M., et al. (2019). Absolute Prioritization of Planetary Protection, Safety, and Avoiding Imperialism in All Future Science Missions: A Policy Perspective. *Space Policy*, *51*, 101345.

Vidaurri, M. and A. Gilbert (2020) "Environmental Considerations in the Age of Space Exploration: The Conservation and Protection of Non-Earth Environments." White paper submitted to this Decadal Survey.

Wiegert, P. (2020). On the delivery of DART-ejected material from asteroid (65803) Didymos to Earth. *The Planetary Science Journal*, *1*, 3.